\documentclass[preprint,showpacs,preprintnumbers,amsmath,amssymb]{revtex4}
\usepackage{graphicx}
\usepackage{dcolumn}
\usepackage{bm}
\usepackage{hyperref}

\begin{document}
\preprint{}
\title{Torsion induces Gravity}
\author{Rodrigo Aros}
\affiliation{Departamento de Ciencias F\'{\i}sicas, Universidad Andr\'es Bello, Av. Republica 252,
Santiago,Chile}
\author{Mauricio Contreras}
\affiliation{Facultad de ciencias y tecnolog\'{\i}a Universidad Adolfo Iba\~nez, Avenida Diagonal las
Torres 2640, Pe\~nalol\'en. Santiago, Chile.}

\date{\today}
\begin{abstract}
In this work the Poincar\'{e}-Chern Simons and Anti de Sitter Chern Simons gravities are studied. For
both a solution that can be casted as a black hole with manifest torsion is found. Those solutions
resemble Schwarzschild and Schwarzschild-AdS solutions respectively.
\end{abstract}
\maketitle

\section{Introduction}
The theory of gravity can be constructed under the idea of a generic covariant theory of fields
with second order equations of motion for the metric. In four dimensions this restriction almost
univocally leads to the EH action.

The inclusion of fermions requires to introduce a local orthonormal basis (vielbein ) on the
manifold and a connection for the local Lorentz group. If vierbein $e^{a}$ and Lorentz connection
$\omega^{ab}$ are considered as independent fields \cite{Zumino:1985dp} arises first order gravity
(see for instance \cite{Zanelli:2002qm}). Usually in first order gravity the fields are written in
terms of differential forms, for instance in four dimensions the first gravity order EH action
reads
\begin{equation}\label{EH}
    I_{EH} = \int_{\mathcal{M}} R^{ab}\wedge e^{c}\wedge e^{d}\varepsilon_{abcd},
\end{equation}
where
\[
R^{ab}=d\omega^{ab} + \omega^{a}_{\hspace{1ex} c} \wedge \omega^{cb}=
\frac{1}{2}R^{ab}_{\hspace{2ex}cd}e^{c}\wedge e^{d},
\]
being $R^{ab}_{\hspace{2ex} cd}$ the Riemann tensor. $\varepsilon_{abcd}=\pm 1,0$ stands for the
complete antisymmetric symbol. Note that the Lagrangian in Eq.(\ref{EH}) is manifestly invariant
under Lorentz transformations.

The equation of motion yielded by the action (\ref{EH}) are
\begin{eqnarray}
  \delta e^{d} &\rightarrow& R^{ab}\wedge e^{c}\varepsilon_{abcd}=0 \label{EinsteinEquations},\\
  \delta \omega^{ab} &\rightarrow& T^{c}\wedge e^{d}\varepsilon_{abcd}=0\label{TorsionEquations},
\end{eqnarray}
where $T^{a}=de^{a} + \omega^{a}_{\hspace{1ex} b}\wedge
e^{b}=\frac{1}{2}T^{a}_{\hspace{1ex}bc}e^{b}\wedge e^{c}$ corresponds to the torsion two form with
$T^{a}_{\hspace{1ex}bc}$ the torsion tensor. Note that Eq.(\ref{TorsionEquations}) is an algebraic
equation, with solution $T^{a}_{\hspace{1ex} cd}=0$. Once this is replaced in
Eqs.(\ref{EinsteinEquations}) they become the standard Einstein equations, therefore any solution
of the metric formalism is recovered on-shell by this formulation. From now on the $\wedge$
product will be understood between differential forms.

In higher dimensions the premise of second order equation of motion for the metric do not restrict
the action to EH. In the language of first order gravity there are many sensible theories of
gravities whose Lagrangians can be constructed in terms only of $R^{ab}$ and $e^{a}$. These
theories of gravity are known as Lovelock gravities \cite{Lovelock:1971yv} and by construction are
manifestly invariant under Lorentz transformations\cite{Zumino:1985dp,Zanelli:2002qm}. Although
the equations of motion do
 not force the vanishing of torsion, the consistency of them,
\textit{e.g.} their covariant derivatives, in general introduce over-determinant conditions unless
torsion vanishes \cite{Zanelli:2002qm}.

However in odd dimensions there is a subfamily of Lovelock gravities whose consistency conditions
do not introduce new conditions to be satisfied, thus in general non vanishing torsion solutions
are permitted. These are known as Chern Simons gravities because they coincide with a Chern Simons
gauge theory, respectively, for the Poincar\'{e} group for a vanishing cosmological constant
($\Lambda=0$), and for the (A)dS group if $\Lambda\neq 0$ \cite{Chamseddine:1989nu}. The
connection of that gauge of theory is constructed as
\begin{equation}\label{connection}
 A^{AB} \sim \left[ \begin{array}{cc}
        \omega^{ab} & e^{a} \\
      -e^{b}& 0 \end{array}\right].
\end{equation}

Unfortunately only a few solutions of Chern Simons gravity are known. For $\Lambda=0$ and
$T^{a}=0$ the solutions are merely flat spaces with angular defects. For $\Lambda<0$ and $T^{a}=0$
black hole solutions are known \cite{Aminneborg:1996iz,Banados:1994ur}. The solutions in Ref.
\cite{Aminneborg:1996iz} have constant curvature, and are obtained through identification of the
AdS space. On the other hand, the solution in Ref.\cite{Banados:1994ur} in $2n+1$ dimensions reads
\[
ds^2 = - F(r) dt^2 + \frac{dr^2}{F(r)} + r^2 d\Omega^{2}_{2n-1}
\]
with $F(r)=1-(M+1)^{\frac{1}{n}}+\frac{r^2}{l^2}$. One can check that this solution is not a
perturbation of the $2n+1$ dimensional Schwarzschild-AdS solution.

In the sections below is shown that remarkably the presence of torsion in the five dimensional
Chern Simons gravity, as an example for higher dimensions, allows solutions that resemble the
Schwarzschild solution with either vanishing or negative cosmological constant. In addition in
section ({\ref{Geodesic}}) is argued that these solutions are indeed black holes. This work
corresponds to a first step to study the role of torsion in (Chern Simons) gravity.

Finally it is worth to stress that in other context, usually called teleparallel gravity, the role
of torsion as an inducer of gravity is well known \cite{Hayashi:1979qx}. This approach, as seen
below, is essentially different.

\subsection{The space}

The spaces in this work, are topologically cylinders. In general $\mathcal{M}=\mathbb{R}\times
\Sigma$ where $\Sigma$ corresponds to a 3-dimensional spacelike hypersurface and $\mathbb{R}$
stands for the time direction. In addition, $\partial\Sigma$ will be considered the union of an
exterior and an interior surface, thus $\partial \Sigma=\partial \Sigma_{\infty} \oplus \partial
\Sigma_{H}$.

\section{$\Lambda =0$ Chern Simons Gravity}
The Poincar\'{e} Chern Simons action in five dimensions reads
\begin{equation}\label{CS}
    I = \int_{\mathcal{M}} R^{ab}R^{cd}e^{f}\varepsilon_{abcdf}.
\end{equation}
yielding the two set of equations of motion
\begin{equation}\label{CS1}
R^{ab}R^{cd}\varepsilon_{abcdf} = 0 \textrm{ and  }R^{ab}T^{c}\varepsilon_{abcdf} = 0.
\end{equation}

\subsection{Spherical symmetric solution}

Considering a spherical symmetric solution one is led to the ansatz
\begin{subequations}\label{Ansatz}
\begin{equation}
\label{vielbein} e^{0}=f(r) dt\textrm{  } e^{1}= g(r)dr\textrm{  }e^{m} = r \tilde{e}^{m},
\end{equation}
\begin{eqnarray}
   \omega^{01}= c(r) dt & & \omega^{mn} =  \tilde{\omega}^{mn},\nonumber\\
  \omega^{0m}= a(r)\tilde{e}^{m} & & \omega^{1m}= b(r)\tilde{e}^{m},\label{spinconnection}
\end{eqnarray}
\end{subequations}
where $\tilde{e}^{m}$ is a dreivein (with $m=2,3,4$) for the spherical transverse section and
$\tilde{\omega}^{mn}$ is its associated Levi Civita (torsion free) connection.

In this ansatz the curvature tensor reads
\begin{eqnarray}
R^{01}&=& -\frac{1}{g(r)f(r)}\frac{dc(r)}{dr} e^{0}e^{1} \nonumber  \\
R^{0m}&=& b(r)\frac{c(r)}{f(r)r}  e^{0}e^{m} + \frac{1}{r g(r)}\frac{da(r)}{dr} e^{1} e^{m}  \nonumber\\
R^{1m} &=& \frac{1}{r g(r)}\frac{db(r)}{dr} e^{1} e^{m} + a(r) \frac{c(r)}{f(r)r}  e^{0}e^{m}  \nonumber  \\
R^{mn} &=& (1+ a(r)^2-b(r)^2)\tilde{e}^{m}\tilde{e}^{n}.\label{Curvatures}
\end{eqnarray}

Torsion tensor on the other hand reads
\begin{eqnarray}
T^{0}&=& \frac{1}{g(r)f(r)}\left(c(r)g(r)-\frac{df(r)}{dr}\right) e^{0}e^{1},  \nonumber\\
T^{m}&=&\frac{1}{r g(r)}\left(1+b(r)g(r)\right)e^{1}e^{m}-\frac{a(r)}{r} e^{0}e^{m}.
\label{Torsions}
\end{eqnarray}

\subsection{Boundary conditions}\label{BoundaryConditionsFlat}

At the asymptotical region $\partial \Sigma_{\infty}$  the standard condition that $\mathcal{M}$
be asymptotically flat will be imposed, \textit{i.e.},
\[ \lim_{x \rightarrow \partial
\Sigma_{\infty}} R^{ab} \rightarrow  0 \textrm{ and } \lim_{x \rightarrow \partial
\Sigma_{\infty}} T^{a}  \rightarrow  0,
\]
is imposed. This in turn determines the behavior of the functions in Eq.(\ref{Ansatz})
\begin{equation}\label{Conditions}
    \begin{array}{cc}
f(r)_{x \rightarrow \partial \Sigma_{\infty}} \rightarrow  1, & c(r)_{x \rightarrow \partial
\Sigma_{\infty}}
\rightarrow  0,\\
g(r)_{x \rightarrow \partial \Sigma_{\infty}} \rightarrow  1, &
a(r)_{x \rightarrow \partial \Sigma_{\infty}} \rightarrow  0,\\
& b(r)_{x \rightarrow \partial \Sigma_{\infty}} \rightarrow  -1.    \end{array}
\end{equation}

Looking for a solution that resembles a black hole solution, and recalling the Schwarzschild
solution, it is also imposed that $f(r)=1/g(r)$ has a simple zero at certain value $r=r_{+}$ being
positive for $r>r_{+}$. This condition in manifold with vanishing torsion would determine the
presence of a Killing horizon at $r=r_{+}$, in this case, nonetheless, this must be proven. The
boundary condition at $r\rightarrow r_{+}$ is regularity of $R^{ab}$.

\subsection{Solution}

The equation of motion (\ref{CS1}), with the ansatz in Eq.(\ref{Ansatz}), implies
$a(r)^2=b(r)^2-1$ with $b(r)^2 > 1$ since $a(r)$ must be real.

Next, imposing the boundary conditions in section (\ref{BoundaryConditionsFlat}) arises the
solution
\begin{eqnarray}
  f(r) = \sqrt{1-\frac{r_{+}^{2}}{r^{2}}}, &&  g(r) =  f(r)^{-1}, \label{solutionExplicit} \\
  c(r) = \frac{r^{2}_{+}}{r^{3}}-\frac{r^{3}_{+}}{r^{4}},& &
  b(r) = -f \left( r \right) + b_{2}\left( r \right)\nonumber,
\end{eqnarray}
where
\begin{eqnarray}
 b_{2}(r)&=& F(r) \left(C_{1} + \int^{r}_{r_{+}} d\rho \frac{1}{F(\rho)}
\frac{d}{d\rho}f(\rho) \right),\nonumber\\
  F(r) &=& r{e^{{\frac {r_{+}}{r}}}} \left( r+r_{+} \right) ^{-1}\nonumber.
\end{eqnarray}

It must be stressed that the vanishing of the torsion at $\partial \Sigma_{\infty}$ determines
that
\[C_{1}=-\int^{\infty}_{r=r_{+}} d\rho \frac{1}{F(\rho)}
\frac{d}{d\rho}f(\rho),
\]
which also satisfies the bound $b(r)^2>1$ for $r\in [r_{+}\ldots \infty]$.

In section (\ref{Geodesic}) is sketched a demonstration that this solution can be casted as a
black hole.

\section{$\Lambda <0$ Chern Simons Gravity}

In this section the problem is restudied for $\Lambda <0$. In five dimensions the AdS Chern Simons
Lagrangian reads
\begin{equation}\label{CSn}
\mathbf{L} = \left(R^{ab}R^{cd}e^{f}+\frac{2}{3l^{2}}
R^{ab}e^{c}e^{d}e^{f}+\frac{1}{5l^{4}}e^{a}e^{b}e^{c}e^{d}e^{f}\right)\varepsilon_{abcdf},
\end{equation}
where the cosmological constant $\Lambda=-6 l^{-2}$. This Lagrangian (\ref{CSn}) generates the two
set of equations
\begin{equation} \label{CS1n}
 \bar{R}^{ab}\bar{R}^{cd}\varepsilon_{abcdf} = 0 \textrm{ and } \bar{R}^{ab}T^{c}\varepsilon_{abcdf} =
 0,
\end{equation}
where $\bar{R}^{ab}=R^{ab}+l^{-2} e^{a}e^{b}$.

\subsection{Boundary conditions and a solution}

To obtain an explicit solution is still necessary to define boundary conditions. In this case the
negative cosmological constant suggests to impose that $\mathcal{M}$ be asymptotically AdS, i.e.,
\[ \lim_{x \rightarrow \partial \Sigma_{\infty}} \bar{R}^{ab}\rightarrow  0
\textrm{ and } \lim_{x \rightarrow \partial \Sigma_{\infty}} T^{a} \rightarrow  0.
\]
Now, in term of the functions in the ansatz (\ref{Ansatz}), this condition determines
\begin{equation}\label{ConditionsNegative}
    \begin{array}{cc}
f(r)_{x \rightarrow \partial \Sigma_{\infty}} \rightarrow  \frac{r}{l}, & c(r)_{x \rightarrow
\partial \Sigma_{\infty}} \rightarrow  \frac{r}{l}, \\
g(r)_{x \rightarrow \partial \Sigma_{\infty}} \rightarrow  \frac{l}{r}, &
a(r)_{x \rightarrow \partial \Sigma_{\infty}} \rightarrow  0,\\
& b(r)_{x \rightarrow \partial \Sigma_{\infty}} \rightarrow  -\frac{r}{l}.
\end{array}
\end{equation}

As in $\Lambda=0$ here the presence of a \textit{horizon} will be introduced through imposing that
$f(r)=1/g(r)$ has at least a zero, called $r_{+}$. The boundary condition as $r\rightarrow r_{+}$
is regularity of $\bar{R}^{ab}$.

Using the spherical symmetric ansatz in Eqs.(\ref{Ansatz}) and imposing
\[
a(r)^2 = b(r)^2- 1 - \frac{r^{2}}{l^{2}},
\]
the equations of motion (\ref{CS1n}) can be solved.  After considering these boundary conditions
one obtains the Schwarzschild like solution
\begin{eqnarray}
f(r)^{2} &=& 1-\frac{\mu}{r^2}+\frac{r^{2}}{l^{2}}\nonumber\\
         &=& \frac{(r-r_{+})(r+r_{+})}{r^{2}}\left(1+\frac{r^{2}}{l^{2}}+\frac{r^{2}_{+}}{l^{2}}\right)\nonumber\\
g(r) &=&  f(r)^{-1}\label{solutionExplicitNegative} \\
c(r) &=&  \frac{r_{+}^{2}}{r^{3}}\left(1+\frac{r_{+}^2}{l^2}\right)+ \frac{r}{l^{2}}-\frac{1}{r}\left(1+\frac{2r_{+}^2}{l^2}\right)\nonumber\\
b(r)&=&-f \left( r \right) + b_{2}\left( r \right)\nonumber
\end{eqnarray}
with
\begin{equation}\label{b2n}
b_{2}(r)= F(r) \left( C_{1} + \int^{r}_{r_{+}} d\rho \frac{1}{F(\rho)}\left[ \frac{d}{d\rho}
f(\rho)-\frac{\rho}{l^{2}f(\rho)}\right] \right)
\end{equation}
where
\[
 C_{1} < -\sqrt{1+\frac{r_{+}^{2}}{l^{2}}}\left(\frac{2r^{2}_{+}+l^{2}}{r_{+}l}\right),
\]
and
\begin{equation}\label{FRnegative}
F(r) = \frac{rl}{r^{2}+r^{2}_{+}+l^{2}}.
\end{equation}

Analogous to the mass, $C_{1}$ is only bounded by the boundary conditions (the existence of a
horizon). The vanishing of $b_{2}(r)$ at $\Sigma_{\infty}$ is trivially satisfied because $F(r)$
vanishes at the spatial infinity. $\Lambda<0$ seems to constrain, unlike $\Lambda=0$, the torsion
to vanish without further constraints on the integration constants.

\section{A horizon}\label{Geodesic}

To confirm the presence of a horizon at $r=r_{+}$ in this case is enough to analyze the behavior
near $r=r_{+}$, therefore one can study both solutions on a general ground.

The presence of a horizon on a torsionless manifold can be uncovered by analyzing the geodesic
curves. In general one can argue that $r=r_{+}$ is a genuine horizon, roughly speaking, only if
there no light like outward curves connecting $r<r_{+}$ to $r>r_{+}$. To realize that in this case
one must recall that in any manifold there are two parallel ways to address the movement of
particles, the straightest and the shortest (or longest) curves, this last known also as the
geodesic curve \cite{Nakahara:1990th}. Unfortunately in a manifold with torsion both definitions
disagree. The tangent vector of the geodesic curves satisfies
\[
 l_{g}^{\mu}\left(\partial_{\mu} l_{g}^{\nu} + \left\{
 \begin{array}{c}
 \nu \\ \alpha \mu \\
\end{array} \right\}
l_{g}^{\alpha}\right) = 0,
\]
where $\{\}$ is the Christoffel symbol, while the straightest curve satisfies
$l_{s}^{\mu}\left(\partial_{\mu} l_{s}^{\nu}+ \Gamma^{\nu}_{\alpha\mu}l_{s}^{\alpha}  \right)=0$
or explicitly
\begin{equation}\label{straightestpossiblecurve}
l_{s}^{\mu}\left(\partial_{\mu} l_{s}^{\nu} +  \left(\left\{
\begin{array}{c}
 \nu \\ \alpha \mu \\
\end{array}
\right\}+ \frac{1}{2} \left(T^{\hspace{1ex}\nu}_{\mu\hspace{1ex}\alpha}+
T^{\hspace{1ex}\nu}_{\alpha\hspace{1ex}\mu} \right)\right) l_{s}^{\alpha}\right) = 0.
\end{equation}

Note that the straightest curve satisfies a non homogenous geodesic equation in a term
proportional to the torsion tensor. Sometimes Eq.(\ref{straightestpossiblecurve}) is interpreted
as a geodesic equation with an effective electric field, due to torsion, although this is only
apparent \cite{DeChingChern}.

To proceed one can restrict the analysis only to the radial curves, \textit{i.e.} the $(t-r)$
plane. After solving both equations the tangent vector of the light like radial geodesic curves
reads
\[
 l_{g} = \frac{1}{f(r)^2} \frac{\partial}{\partial t}\pm  \frac{\partial}{\partial r},
\]
while the tangent vector of the light like radial straightest curves is
\begin{equation}\label{Straightest}
l_{s} = e^{-\int \frac{c(r)}{f^{2}}}\left(\frac{1}{f(r)} \frac{\partial}{\partial t}\pm  f(r)
\frac{\partial}{\partial r} \right)= (f(r)e^{-\int \frac{c(r)}{f^{2}}}) l_{g}.
\end{equation}
Since the tangent vectors are parallel one can show that both curve are equivalent modulo a
redefinition of the affine parameter.

The timelike curves, on the other hand, are not equivalent. Nonetheless the boundary conditions
ensure that as $r\rightarrow \infty$ both curves must converge.

Since radial light like curves are equivalent and $l_{g}$ depends only on the metric $r=r_{+}$ is
a horizon in the sense that $r=r_{+}$ is a light like surface. This result grants that the name
\textit{black hole} can be applied to the solutions presented in this work.

To determine additional characteristics of the solutions probably one needs to study fields
evolving on them, if not the back reaction too. The integer spin fields couples gravity only
through the vielbein (or the metric) at tree level, but at higher loops, however, the presence of
torsion affects for instance the propagation of the electromagnetic fields
\cite{deSabbata:1994wi}. Half spin fields, on the other hand, couple through the spin connection,
and so they couple the torsion tensor. Thus, fields can be useful to unveil characteristics of the
manifold, however although this is an interesting direction to continue, is beyond the scope of
this work.

\section{Conclusions and discussion}

In this work is proven that Chern Simons gravity in five dimensions has solutions that resemble
Schwarzschild black holes. These solutions have non vanishing torsion, but the standard geodesic
structure of black hole is preserved, though. In both solutions
(Eqs.(\ref{solutionExplicit},\ref{solutionExplicitNegative})) torsion vanishes fast enough, in
particular for $\Lambda <0$, that its influence can be neglected in a region still far from
$\mathbb{R}\times\partial\Sigma_{\infty}$, thus  both solutions represent interesting
generalizations of Schwarzschild solution, preserving its general behavior.

One can expect the results in this work can be readily extended to higher odd dimensions, thus
Schwarzschild like black holes with torsion should exist in any odd dimensions within Chern Simons
gravity. This can be particular interesting in 11 dimensions, where a supergravity theory in terms
a Chern Simons action is known \cite{Banados:1996hi}.

Chern Simons theories, and in particular Chern Simons gravities, are known to have complicated
phase spaces \cite{Banados:1996yj,Miskovic:2005di}. These solutions probably can be a contribution
to understand the role of torsion in Chern Simons gravities, in particular because the presence of
torsion reveals degrees of freedom usually ignored in the torsion free solutions. Associated with
that, one controversial issue of Chern Simons gravity is the possibility of changing the geometry,
maybe even the topology, of a solution through a gauge transformation. Because of that one usually
speaks of permitted and non permitted gauge transformations. For instance a gauge transformation
which nullifies the part of the connection corresponding to the vielbein (See
Eq.(\ref{connection})) must be forbidden. Analogously, for Chern Simons gravity the curvature and
the torsion are part of a larger field strength, thus in principle there is a group element $g$
such that
\[
  F= \left[\begin{array}{cc}
       R^{ab} & T^{a} \\
       -T^{a} & 0
        \end{array}\right] \mapsto \tilde{F }=  g^{-1} F g=\left[\begin{array}{cc}
       \tilde{R}^{ab} & 0 \\
       0& 0
        \end{array}\right],
\]
relating torsion free spaces to the solutions
(\ref{solutionExplicit},\ref{solutionExplicitNegative}). For the two solutions under study,
although it is direct to prove that the transformation exists, it is also direct to prove that the
related solutions are not known solutions, neither the Schwarzschild solution nor the solution in
\cite{Banados:1994ur}. Actually the related solution are not a static black holes.

\begin{acknowledgments}
R.A would like to thanks O. Chandia, A. Gomberoff and R. Herrera for heplful conversations. R.A
would like to thank Abdus Salam International Centre for Theoretical Physics (ICTP) for the
associate award granted. This work was partially funded by grants FONDECYT 1040202 and DI 06-04.
(UNAB).
\end{acknowledgments}


\providecommand{\href}[2]{#2}\begingroup\raggedright\endgroup

\end{document}